\documentclass[twocolumn,aps,showpacs,prl,superscriptaddress]{revtex4}%,showpacs]{revtex4}
\usepackage{graphicx}
\usepackage{dcolumn}% Align table columns on decimal point
\usepackage{bm}% bold math
 \usepackage{amssymb}

\begin{document}

\title{A new collective phenomenon arising from spin anisotropic perturbations to a Heisenberg square lattice manifested in paramagnetic resonance experiments }  

\author{S. Cox}
\affiliation{National High Magnetic Field Laboratory, Los Alamos National Laboratory, MS-E536, Los Alamos, NM 87545, USA}
\author{R.D. McDonald}\email{rmcd@lanl.gov}
\affiliation{National High Magnetic Field Laboratory, Los Alamos National Laboratory, MS-E536, Los Alamos, NM 87545, USA}
\author{J. Singleton} 
\affiliation{National High Magnetic Field Laboratory, Los Alamos National Laboratory, MS-E536, Los Alamos, NM 87545, USA}
\author{S. Miller} 
\affiliation{National High Magnetic Field Laboratory, Los Alamos National Laboratory, MS-E536, Los Alamos, NM 87545, USA}
\author{P.A. Goddard}
\affiliation{Clarendon Laboratory, Department of Physics, Oxford University, Oxford, UK OX1 3PU}
\author{S. El Shawish}
\affiliation{J. Stefan Institute, Jamova cesta 39, 1000 Ljubljana, Slovenia}
\author{J. Bonca}
\affiliation{Faculty of Mathematics and Physics, University of Ljubljana, Jadranska, 1000 Ljubljana, Slovenia}
\affiliation{J. Stefan Institute, Jamova cesta 39, 1000 Ljubljana, Slovenia}
\author{J.A. Schlueter}
\affiliation{Materials Science Division, Argonne National Laboratory, Argonne, IL 60439 USA}
\author{J.L. Manson} 
\affiliation{Department of Chemistry and Biochemistry, Eastern Washington University, Cheney, WA 99004, USA}

\begin{abstract}

We report unexpected behaviour in a family of Cu spin-$\frac{1}{2}$ systems, in which an apparent gap in the low energy magneto-optical absorption spectrum opens at low temperature. This previously-unreported collective phenomenon arises at temperatures where the energy of the dominant exchange interaction exceeds the thermal energy. Simulations of the observed shifts in electron paramagnetic resonance spectral weight, which include spin anisotropy, reproduce this behavior yielding the magnitude of the spin anisotropy in these compounds.
\end{abstract}


\maketitle

The spin ($S$) $\frac{1}{2}$ two-dimensional (2D) 
square-lattice quantum Heisenberg 
antiferromagnet system has long been interesting 
to theoretical physicists due to the variety of transitions that can arise~\cite{heis_model_1,heis_model_2,manousakis}. 
Moreover, the role of $S=\frac{1}{2}$ fluctuations on a 
square lattice in the mechanism for cuprate superconductivity
is hotly debated~\cite{manousakis,cuprate_book,julian_cuprates,neil_mad_article}.
The recently discovered family of H-bonded metal-organic 
magnets~\cite{manson_chemcomm} 
offer the possibility to readily control the exchange 
parameters in a 2D system by changing chemical composition, 
thus creating spin architectures with desirable properties `to order'~\cite{goddard}. 
For an idealized 2D system, long range magnetic order would not occur at finite temperature~\cite{heis_model_2,Neelzero}.  However, in the metal-organic systems, 
interlayer coupling gives rise to a finite Neel 
temperature~\cite{Neelfinite1,Neelfinite2,Neelfinite3}. 
For these quasi-2D systems the ordering temperature is 
dominated by the weakest (the interlayer) exchange interaction, 
whereas the saturation magnetic field is dominated by the 
strongest exchange interactions, thus providing a means of 
estimating the spatial exchange anisotropy in the system~\cite{goddard}. 
It should be noted that the more 2D the system, 
the wider the temperature ($T$) range, $T_{\rm N} < T < J/k_{\rm B} $, 
over which magnetic fluctuations dominate. 
Here we demonstrate that a spin anisotropy perturbation to 
the Heisenberg square lattice results in a new collective 
phenomenon within this regime, manifested as a shift in electron 
paramagnetic resonance (EPR) frequency at low temperature.

\begin{figure}
\centering
\includegraphics [width=0.85\columnwidth]{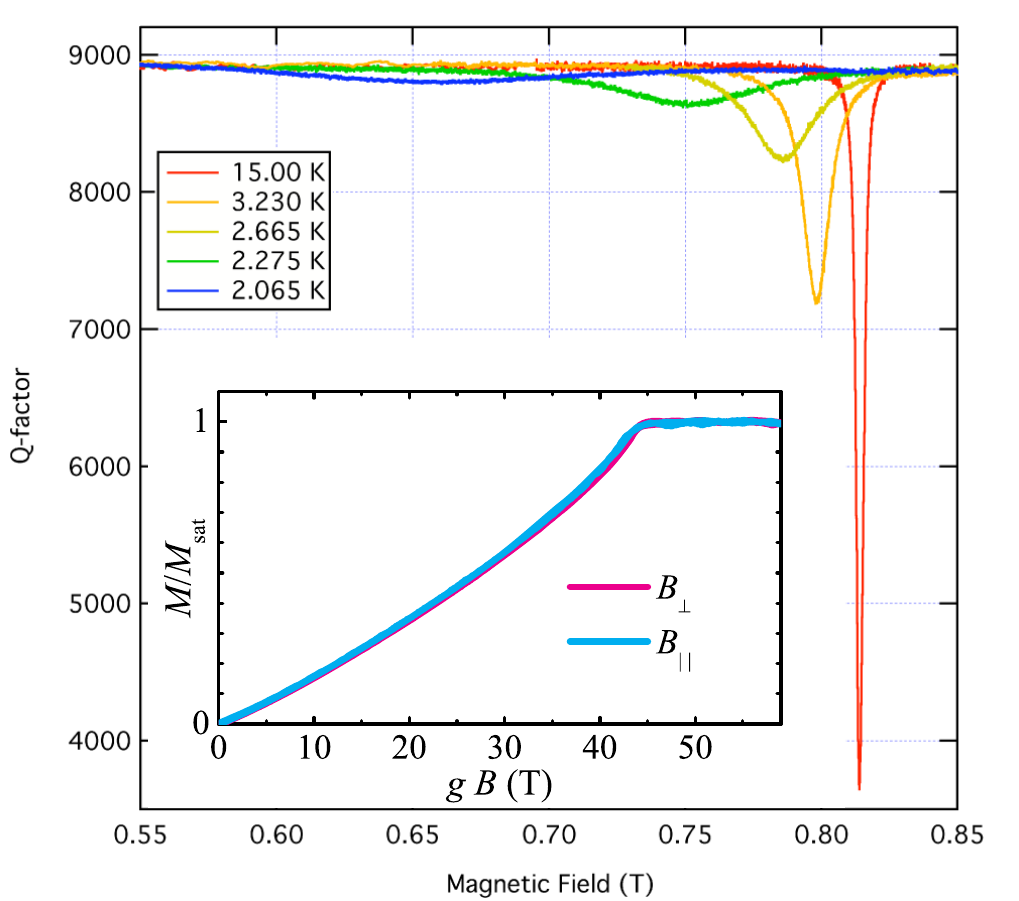}
\caption{a) The Q-factor of the 26~GHz cavity-mode as a function of magnetic field, illustrating the evolution of the EPR line width and resonant magnetic field with temperature in [CuHF$_2$(pyz)$_2$]ClO$_4$. The data is for magnetic field applied perpendicular to the planes. (b) Low-temperature magnetisation data in the same compound~\cite{goddard}.}
\label{Fig1}
\end{figure}

Fig~\ref{Fig1} illustrates the dramatic shift of the EPR
magnetic field and line width with temperature for a single frequency, $f$. 
Multi-frequency EPR measurements, fitted to $f = f_0 + g \mu_{\rm B} B/h$,
where $B$ is the resonance field, show that the shifts correspond to changes
in {\it both} the low-field intercept $f_0$ and effective g-factor $g$.
At high $T$, $f_0 =0$, but as the phenomenon
develops with decreasing $T$, $f_0$ becomes finite and positive,
behavior that strongly resembles an energy gap opening.
Not only is this unexpected in an $S=\frac{1}{2}$ system;
it is ruled out by the low temperature magnetization,
which shows a monotonic increase between $H=0$ and the 
onset of saturation (see Fig.~\ref{Fig1}).
Any gap would be manifested as a region of reduced or 
zero d$M$/d$H$~\cite{goddard}.

We shall show below that this dramatic change is not linked to the 
antiferromagnetic transition temperature in either of the compounds studied,
but is rather related to the intralayer exchange energy $J$. 
The metal-organic compounds [CuHF$_2$(pyz)$_2$]X 
were therefore selected for this study as the choice of 
anion molecule X can alter $J$ controllably by a factor $\sim 2$
and because the $J$ values have been determined to a high
accuracy using magnetometry~\cite{goddard}.
In these compounds,  Cu$^{2+}$ ions are arranged in square-lattice layers,
separated by pyrazine (pyz) molecules. 
The layers are held apart by bifluoride 
bridges~\cite{manson_chemcomm, Neelfinite3, goddard}.
Single crystals of [CuHF$_2$(pyz)$_2$]X with X = ClO$_4$ or PF$_6$,
were produced by an aqueous chemical reaction between the appropriate 
CuX$_2$ salt and stoichiometric amounts of the ligands (see~\cite{manson_chemcomm, Manson_preprint, Lancaster_PRL07}
for preparation method details and X-ray data).
Both materials undergo antiferromagnetic ordering, with 
$T_{\rm N}=1.94$~K for  X = ClO$_4$ 
and $T_{\rm N}=4.31$~K for X = PF$_6$.
Note that owing to differing summing conventions the definition of $J$ used in
the current paper is a factor two smaller than that in Ref.~\cite{goddard}, {\it i.e.} herein $J_{\rm plane}$(X = ClO$_4$) = 3.6~K  and $J_{\rm plane}$(X = PF$_6)$ = 6.2~K.

EPR spectra of single-crystal samples were obtained in two ways.
In the first method, the sample was placed in a cylindrical resonant cavity 
and the Q-factor and resonant frequency of the cavity 
were measured at each field point.  The change in Q-factor is 
proportional to the microwave absorption of the sample~\cite{abragamandbleaney}.
These measurements were carried out in
the frequency range 11-40~GHz using a Hewlett-Packard 8722ET network analyzer. 
The cavity was placed in a $^4$He flow cryostat that 
was capable of stabilizing temperatures down to 1.5~K.
For higher frequencies, the sample was placed in a confocal 
etalon resonator. The transmission, and hence change in microwave 
absorption, was measured in the frequency range 60-120~GHz 
using an ABmm~\cite{Rossrevsci} Millimetre wave Vector Network 
Analyser. The etalon was placed in a $^3$He system that 
was capable of stabilizing temperatures down to 0.6~K.
For fixed temperatures, the EPR field was found to
be a linear function of frequency, allowing the
intercept and effective g-factor to be determined as
described above; experimental values will
be compared with theoretical predictions below.
\begin{figure}
\centering
\includegraphics [width=0.85\columnwidth]{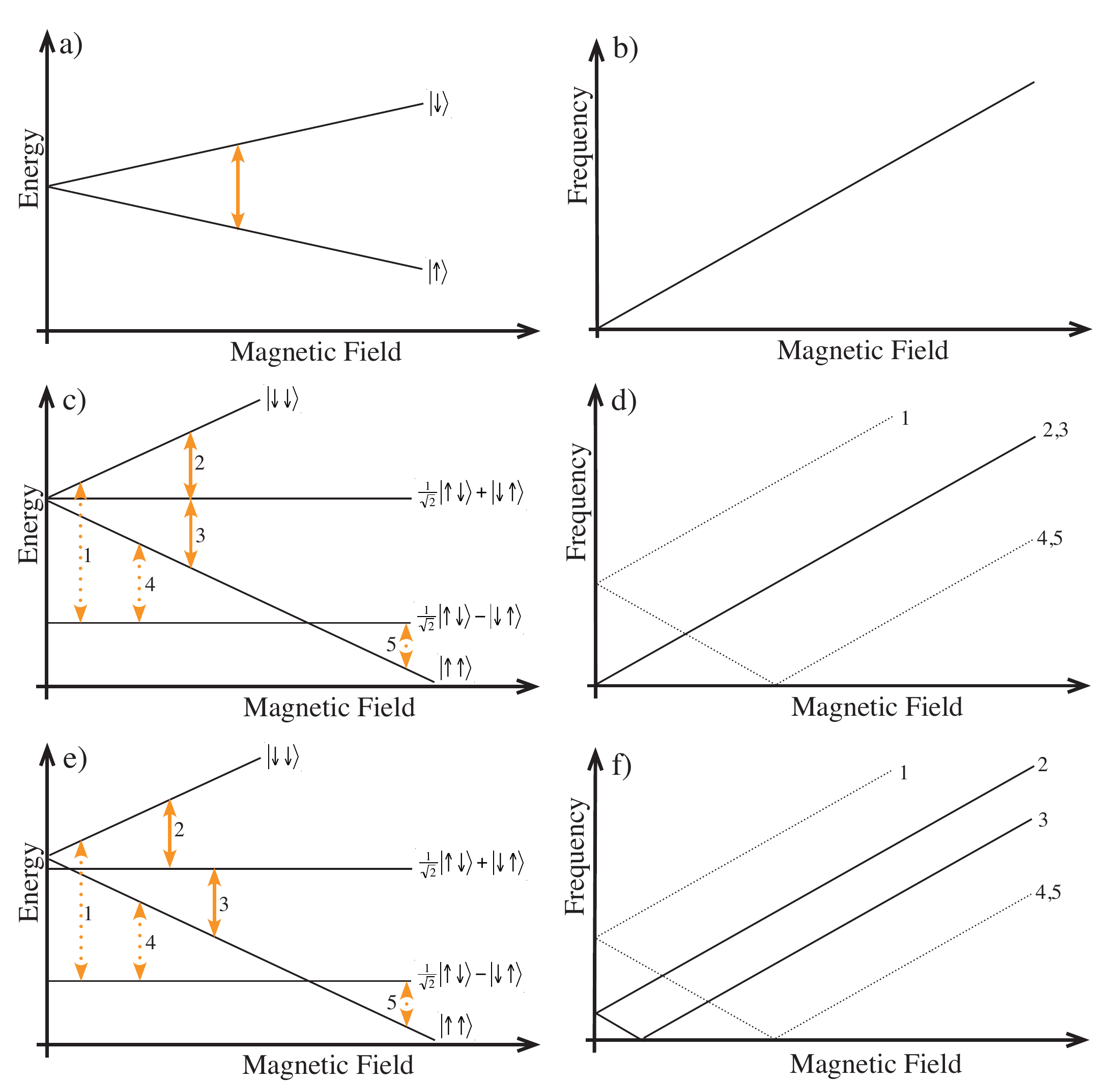}
\caption{Illustration of how a spin-anisotropic exchange interaction can give rise to a an EPR spectrum with a finite frequency intercept in the case of an antiferromagnetically coupled dimer. Respectively (a), (c) and (e) show the Zeeman splitting for a non-interacting, isotropically coupled and spin-anisotropically coupled dimer. (b), (d) and (f) illustrate the corresponding  frequency-magnetic field relationships of the resonant absorption. }
\label{Fig2}
\end{figure}

Before demonstrating that the shift in intercept $f_0$ and effective g-factor $g$
(Fig.~1) is a collective phenomenon involving a large number of spins, 
we show that simpler models such as dimerization cannot account for the data. 
Fig.~\ref{Fig2}a) illustrates the Zeeman splitting for non-interacting electrons and b) the corresponding linear frequency-magnetic field relationship with a zero frequency intercept~\cite{abragamandbleaney}.  
Fig.~\ref{Fig2}c) and d) includes the effect of an antiferromagnetic scalar exchange interaction $J$.
The $|\uparrow \uparrow \rangle$, 
$\frac{1}{\sqrt{2}}(|\uparrow \downarrow \rangle +|\downarrow \uparrow \rangle)$ 
and $|\downarrow \downarrow \rangle$
states will be degenerate at $B=0$, with the 
$\frac{1}{\sqrt{2}}(|\uparrow \downarrow \rangle -|\downarrow \uparrow \rangle)$
state separated by an energy $-J$~\cite{abragamandbleaney}. 
Although this singlet-triplet gap causes the optically active 
($\delta S_z = \pm1, \delta S_{\rm Tot} = 0$) EPR transitions 
within the triplet states to become `frozen out' at 
$T < J$, it does not introduce a gap across which optically active 
EPR transitions occur, which would lead to a finite frequency intercept. 

Fig.~\ref{Fig2}e) and f) includes a spin anisotropy in the exchange interaction, $J_{zz} \neq J_{xx} = J_{yy}$, lifting the zero field degeneracy of the $S_{\rm Tot}=1, S_z=\pm 1$
and $S_{\rm Tot}=1, S_z=0$ states (Fig.~\ref{Fig2}), which 
in turn lifts the field degeneracy of the two intratriplet transitions, 
separating EPR lines 2 and 3.
Although some of the possible EPR lines now 
have finite frequency $B=0$ intercepts,
this simple model does not reproduce the systematics of the effect
reported here.
$\frac{1}{\sqrt{2}}(|\uparrow \downarrow \rangle -|\downarrow \uparrow \rangle)$
is the groundstate so that the two transitions 
potentially observable at low temperature
will be lines 1 and 4; however, for these to be optically-active, 
off-diagonal exchange terms (e.g. $J_{xy}$) such 
as the Dzyaloshinsky-Moriya interaction
must be present (since they mix the singlet and triplet states)~\cite{abragamandbleaney}.
We note that despite a Dzyaloshinsky-Moriya term arising from a lower order spin-orbit perturbation than the diagonal spin anisotropy terms, that they are precluded by the inversion symmetry about the mid point of the dominant exchange interaction. At elevated temperatures $ T > J$, one would observe an EPR spectrum dominated by
lines 2 and 3 (each with a finite $B=0$ intercept); as the sample
cooled to $T < J$, this would change to a spectrum dominated 
by the much weaker (or even absent) lines 1 and 4,
with an interecept differing from that of 2 and 3. 
Such behavior does not lead to a smooth thermal 
evolution of the EPR line or the preservation of spectral 
weight to temperatures $T < J$ that we observe. 

Having shown that a simple local distortion cannot cause the effect
shown in Fig.~1, we now turn to a finite cluster approach applied
to an anisotropic 2D Heisenberg model with spin anisotropy for two field orientations:
\[
H = J \sum \rm{S_i^z S_j^z} + \Delta ({\rm S_i^x S_j^x + S_i^y S_j^y}) - g_{\rm T} \mu_{\rm B} \rm{B} \sum \rm{S_i^{z,y}},
\]
where $\Delta$ is the spin anisotropy, with $\Delta=1.0$ being 
the isotropic (pure Heisenberg) case, $\Delta < 1$ an Ising-like 
or easy-axis antiferromagnet and $\Delta >1$ an easy-plane 
$xy$-antiferromagnet.  It should be noted that the input parameter to the model $g_{\rm T}$ is a temperature independent g-tensor that reproduces the high temperature $(T >> J)$ g-factor anisotropy arising from the spin-orbit interaction. This g-anisotropy is consistent with the magnetic $d_{x^{2}-y^{2}}$ orbital lying in the 2D planes~\cite{goddard}.
 $J_{\rm perp}$ is assumed to be zero, 
since $J_{\rm perp} << J_{\rm plane}$, as demonstrated in~\cite{goddard}.
This model was calculated using  full diagonalization at finite temperature for 12, 14 and 16 sites \footnote{for $B // y$ the maximum number of sites used was 14 due to the lack of $S_z$ symmetry.}.
There was little difference between the results with different numbers of sites, and 
therefore only the 14 site data are displayed here.  The simulations were carried out for 
anisotropy values $\Delta$=1.02, 1.04, 1.06 and 1.08.
It was found that increased spin anisotropy leads to the parallel shift of the peak away 
from its initial position while at the same time the peak gains a finite width.
Multiple smaller peaks in addition to the main peak were 
observed in the simulations for $\Delta \neq 1$, due to finite-size effects.

To compare the simulations to the results for  [CuHF$_2$(pyz)$_2$]ClO$_4$, the 
resonant magnetic field, Lorentz linewidth and spectral weight of the resonance
were calculated for a resonance at 26~GHz.
The resonant magnetic field was calculated as
$B_{\rm R}=2B_0-\langle\omega\rangle$, where $B_0 = 0.35J$ (26~GHz in units of $J$) and
\[
\langle\omega\rangle= \frac{ \int \omega {\rm S}_{xx}(\omega,B_0) \rm{d}\omega}{\int {\rm S}_{xx}(\omega, B_0) \rm{d}\omega}
\]
The Lorentz linewidth 
\[
\Delta B \propto \sqrt{\frac{\langle\omega_{\rm{rel}}^2\rangle^3}{\langle\omega_{\rm{rel}}^4\rangle}}
\]
where $\omega_{\rm{rel}}=\omega-\langle\omega\rangle$.  
The spectral weight was calculated as:
\[
I= \int \xi''(\omega_0,B){\rm d} B \sim (1-{\rm e}^{-\omega_0/T}) \int {\rm S}_{xx}(\omega,B_0)\rm{d}\omega
\]
with the dynamic spin structure factor
\[
{\rm S}_{xx}(\omega) = \frac{1}{N} Re \int_{0}^{\infty}  e^{(i \omega t)} \langle {\rm S}_x(t){\rm S}_x(0)\rangle {\rm d}t.
\]

\begin{figure}
\centering
\includegraphics [width=0.80\columnwidth]{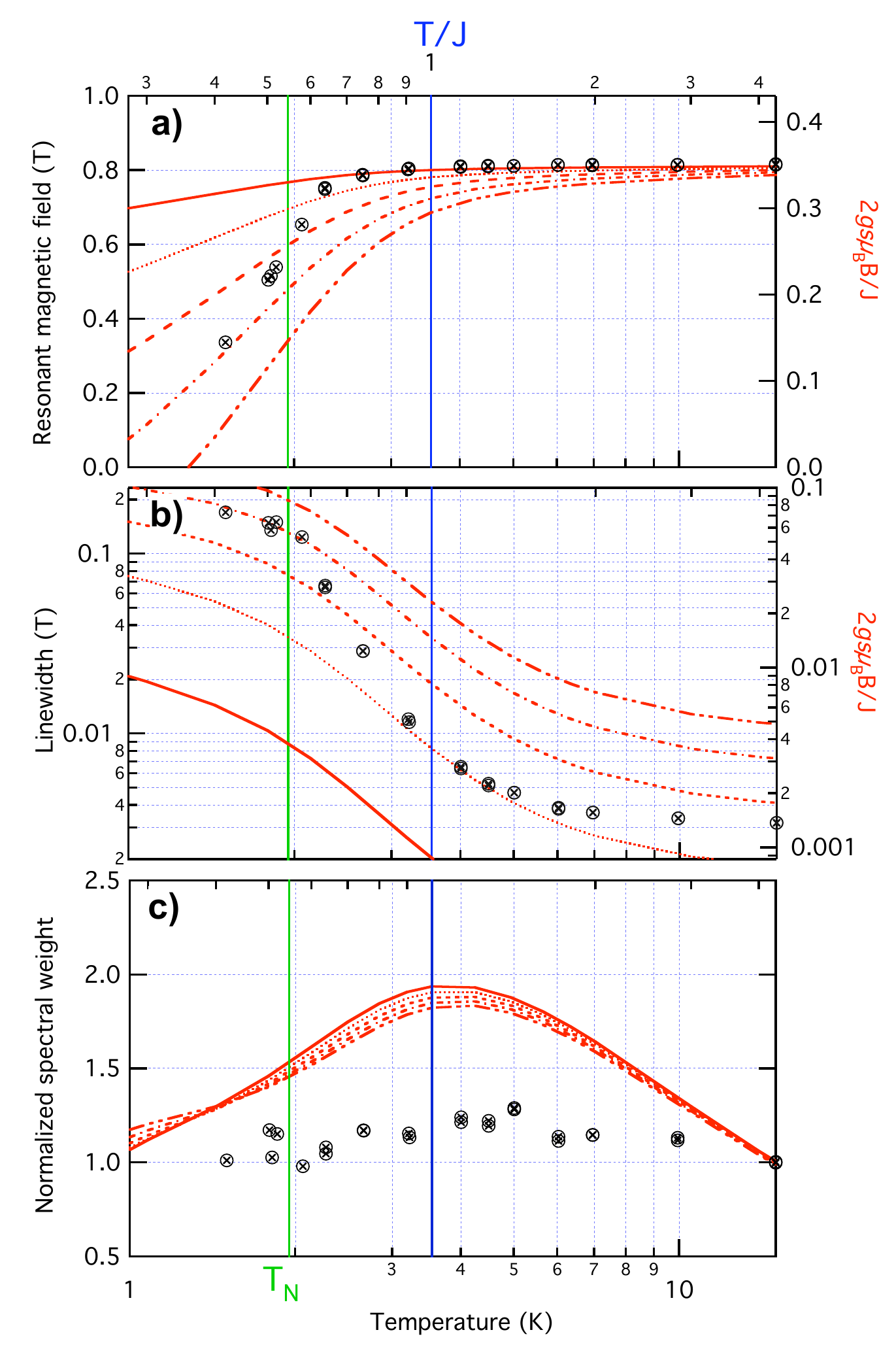}
\caption{A comparison of the experimental 
EPR data, circles, from [CuHF$_2$(pyz)$_2$]ClO$_4$ measured at 26~GHz
with theoretical simulations, $\Delta = $1.02, 1.04, 1.06, 1.08, 1.10; solid, dot, dash, dot-dash and dot-dot-dash respectively. The left and bottom axes are absolute units, the right and top axes are renormalized by the exchange energy and spectroscopic g-factor.  a) The variation of the magnetic field at which resonance occurs as a function of temperature. b) The linewidth (full width at half maximum of a Lorentzian fit) of the resonance peak as a function of temperature. c) The spectral weight (area under Lorentzian fit) normalized to its 15~K value. In all the cases error bars are smaller than data points.}
\label{Fig3}
\end{figure}

A Lorentzian fit to the experimental data yielded the linewidth and spectral weight.
As can be seen from Fig.~\ref{Fig3}, the results of the simulations reproduce the salient features of  
experimental results for [CuHF$_2$(pyz)$_2$]ClO$_4$.
The experimental values clearly lie between the simulated values of 
$\Delta = 1.04$ and 1.08.  Taking into account the fact that above $T=0.8~J$ 
the 1.04 simulation provides a good fit for both 
the position of the resonance and its linewidth, the best match for the data
is given by a spin anisotropy of 1.04.

\begin{figure}
\centering
\includegraphics[width=0.97\columnwidth]{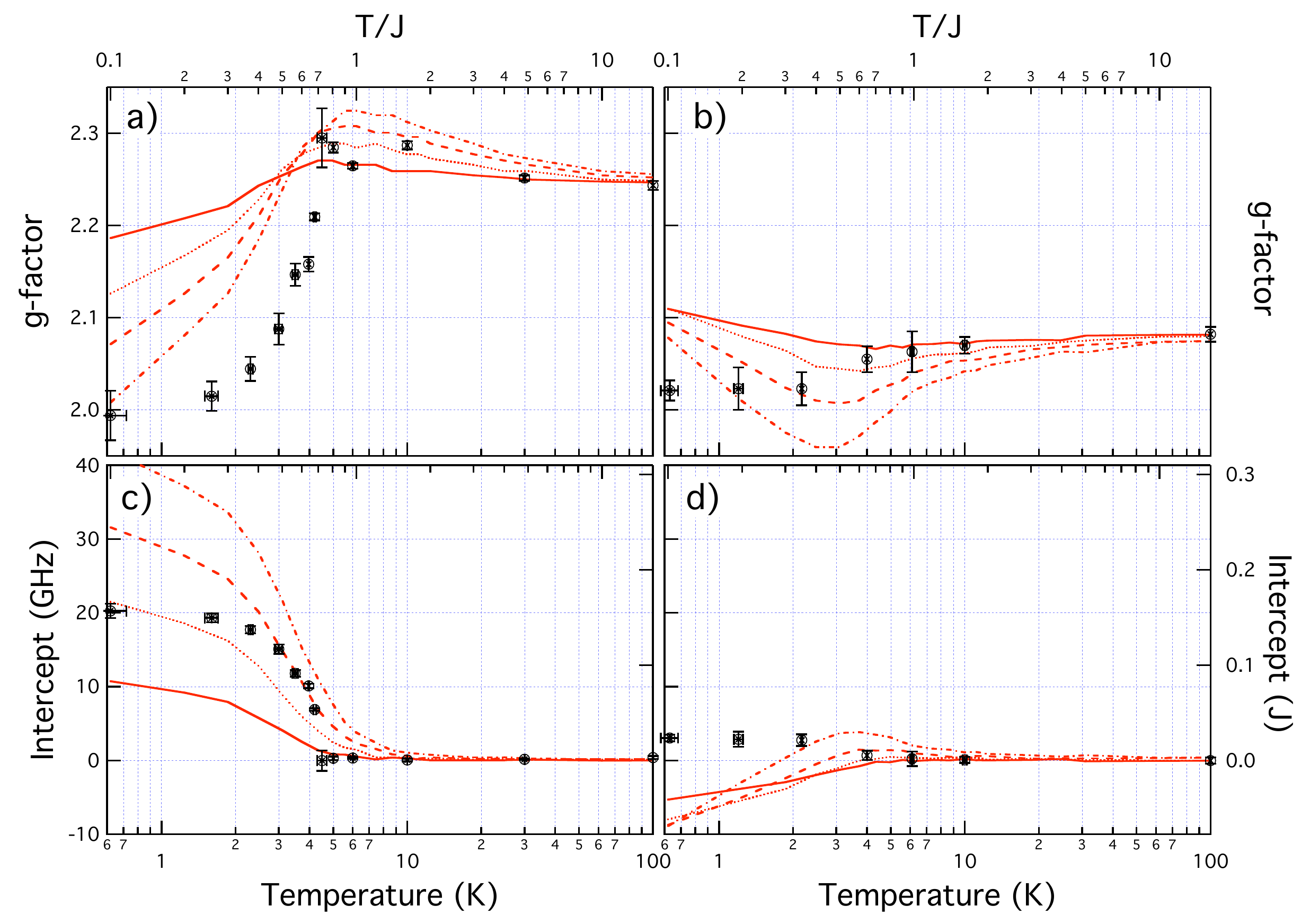}
\caption{Comparison of the experimental g-factor for X = PF$_6$ to simulations for different values of the spin anisotropy, $\Delta = $1.02, 1.04, 1.06, and 1.08, solid, dotted, dashed and dot-dashed lines respectively, for (a) {\bf B} perpendicular to and (b) {\bf B} parallel 
to the Cu square latttice.  Also, comparison of the experimental frequency intercept of resonant frequency vs magnetic field and simulated values for (c) {\bf B} parallel and 
(d) {\bf B} perpendicular.}
\label{Fig4}
\end{figure}

From Fig~\ref{Fig3}  the changes in the position and width of the EPR line
in [CuHF$_2$(pyz)$_2$]ClO$_4$, start as the temperature is lowered
through 3.5~K; this corresponds to $T/J$=1 (as opposed to $T_{\rm N} = 1.94$~K).
Turning to [CuHF$_2$(pyz)$_2$]PF$_6$,
Fig.~\ref{Fig4} 
shows that both $g$ and $f_0$ undergo a dramatic change
on cooling through 4.5~K. Coincidentally, this is close to the ordering 
temperature $T_{\rm N}$~\cite{goddard}.  
The greater relative thermal separation 
of $T_{\rm N}$ and $T = J$ in the ClO$_4$ compound than in the
PF$_6$ salt
reflects the smaller interlayer exchange energy;
one might say that  the former is a closer approximation 
to two dimensionality than the PF$_6$ compound~\cite{goddard}.  
In spite of the proximity of the ordering temperature
to the onset of the EPR shifts, the fit of the model to the 
data from the PF$_6$ compound that we will now give
shows that it is the exchange interaction $J$ that
determines the temperature scale of the effect and not $T_{\rm N}$.

To compare the simulations for the results for [CuHF$_2$(pyz)$_2$]PF$_6$ the 
experimental g-factors and frequency intercepts were calculated.
The theoretical g-factors and intercepts are 
given by the slopes and intercepts of the $\omega, B)$ plots,
which were obtained from the extrapolation through five calculated points,
$\omega/\rm{J}=1.0,1.1,1.2,1.3,1.4$ for correspondence with the experimental frequency range. 

In Fig.~\ref{Fig4} we can see that the behaviour of the experimental g-factor and 
intercept are well matched by the simulations, with the dramatic difference between 
field {\bf B} perpendicular to and {\bf B} parallel to the square Cu plane being reproduced.  
For the {\bf B} parallel data, the fact that we do not observe an upturn in the g-factor and 
the small positive intercept at low temperature is most likely due to a small angular 
misalignment (since the {\bf B} perpendicular effects are so much larger).

Considering the {\bf B} perpendicular data, the g-factor experimental data suggests a spin 
anisotropy above 1.08, whereas the intercept data suggests an anisotropy close to 1.04.
If we consider the corresponding data for X = ClO$_4$ (for which data was
taken at only a few values due to the extreme weakness of the signal at low temperatures) 
we find that at $T/J=0.27$ the experimental values are $ g = 0.96~g_0$ and intercept $= 0.12~J$.
This gives a similar anisotropy, 1.06 - 1.04.
For both materials it can be clearly seen that for $T/J < 1$ 
the experimental changes in both intercept and g factor are 
substantially more rapid than the simulations suggest.  
We propose that this is due to 
the experimental temperature scale being contracted relative to the theoretical temperature scale upon entering the ordered state, {\it i.e.} the phase 
transition has the effect of contracting the entropic (theoretical temperature) scale.  The shape of the intercept curve suggests that the experimental value 
of the intercept has nearly stabilised 
by the lowest $T/J$ values and therefore the 
intercept is taken to give the best
indication of the spin anisotropy (around 1.04 in both compounds).
The approximate conservation of spectral weight in 
both experimental and theoretical results also demonstrates the robustness of our model.

In conclusion we have observed a collective phenomenon in two members of the organic magnet system (Cu(HF$_2$)(pyz)$_2$)X that produces a shift in the frequency intercept of the EPR data which resembles, but does not correspond to, a gap opening in the system. Analogously to an anisotropic g-tensor, the spin anisotropy in the exchange interaction responsible for this effect most likely originates from spin-orbit coupling. As a result, this collective phenomenon is expected to be strongest for low-spin transition metal ions in relatively low symmetry environments, like the octahedral copper site ($3d^9$) in (Cu(HF$_2$)(pyz)$_2$)X. Although the organic systems investigated provide an ideal `low exchange energy scale' environment in which to characterize this effect, strongly coupled copper octahedra are ubiquitous in correlated electron systems; for example, $J \approx 100's$~K in the parent phase of the high $T_{\rm C}$ superconductors.

Work at the NHMFL occurs under the auspices of the National Science Foundation, DoE and the State of Florida. Work
at Argonne is supported by  a U.S. Department of Energy Office of Science laboratory, operated under Contract No. DE-AC02-06CH11357. The authors would like to thank Pinaki Sengupta and Cristian Batista for valuable discussions. 

\bibliographystyle{apsrev}
\bibliography{orgmag_epr7}

\end{document}